\documentclass[a4paper,twoside]{article}

\usepackage{epsfig}
\usepackage{subcaption}
\usepackage{calc}
\usepackage{amssymb}
\usepackage{amstext}
\usepackage{amsmath}
\usepackage{hyperref}
\usepackage{amsthm}
\usepackage{multicol}
\usepackage{pslatex}
\usepackage{apalike}
\usepackage{adjustbox}
\usepackage[dvipsnames]{xcolor}
\usepackage{algorithm2e}
\usepackage[bottom]{footmisc}
\usepackage{SCITEPRESS}     

\newcommand\pernb[1]{\alpha_{#1}}
\newcommand\start[1]{s_{#1}}
\newcommand\horis[1]{u_{#1}}
\newcommand\vertis[1]{v_{#1}}
\newcommand\horip[1]{x_{#1}}
\newcommand\vertip[1]{y_{#1}}
\newcommand\larg{w}
\newcommand\hyperp{H}

\newcommand\rect[1]{R_{#1}}
\newcommand\height[1]{h_{#1}}
\newcommand\base[1]{b_{#1}}
\newcommand\job[1]{J_{#1}}
\newcommand\bbase[1]{B_{#1}}
\newcommand\proctime[1]{p_{#1}}
\newcommand\per[1]{T_{#1}}
\newcommand\pertask[1]{\per{\pernb{#1}}}

\newcommand\flip{flip}
\newcommand\bflip{Bf}
\newcommand\utilization[1]{\eta_{#1}}
\newcommand\utilinst{U}

\newtheorem{lemma}{Lemma}
\newtheorem{theorem}{Theorem}
\newtheorem{definition}{Definition}

\begin{document}
\title{Packing-Inspired Algorithms for Periodic Scheduling Problems with Harmonic Periods}

\author{\authorname{Josef Grus\sup{1,4}, Claire Hanen\sup{2,3} and Zdeněk Hanzálek\sup{4}}
\affiliation{\sup{1}DCE, FEE, Czech Technical University in Prague, Czech Republic}
\affiliation{\sup{2}Sorbonne Universit\'e, CNRS, LIP6, F-75005  Paris, France}
\affiliation{\sup{3} UPL, Universit\'e Paris Nanterre, F-92000 Nanterre, France}
\affiliation{\sup{4}IID, CIIRC, Czech Technical University in Prague, Czech Republic}
\email{josef.grus@cvut.cz; claire.hanen@lip6.fr; zdenek.hanzalek@cvut.cz}}

\keywords{Periodic Scheduling, Harmonic Periods, Height-Divisible 2D Packing, First Fit Heuristic.}
\abstract{We tackle the problem of non-preemptive periodic scheduling with a harmonic set of periods. Problems of this kind arise within domains of periodic manufacturing and maintenance, and also during the design of industrial, automotive, and avionics communication protocols, where efficient scheduling of messages is crucial for the performance of a time-triggered network. We consider the decision variant of the periodic scheduling problem on a single highly-utilized machine. We first prove a bijection between periodic scheduling and a particular (so-called height-divisible) 2D packing of rectangles. We formulate the problem using Constraint Programming and compare it with equivalent state-of-the-art Integer Linear Programming formulation, showing the former's superiority on difficult instances. Furthermore, we develop a packing-inspired first fit heuristic, which we compare with methods described in the literature. We justify our proposed methods on synthetically generated problem instances inspired by the communication of messages on one channel.}

\onecolumn \maketitle \normalsize \setcounter{footnote}{0} \vfill

\section{\uppercase{Introduction and related work}}
\label{sec:introduction}
Periodic Scheduling Problems ($PSPs$) often arise when dealing with manufacturing and maintenance schedules. Another important application area is the configuration of communications protocols in avionics and automotive. The configuration is based upon an offline generated schedule specifying start times of so-called time-triggered messages, typically used for safety-relevant functions, such as drive-by-wire. It is important to have an efficient solution because the demand for communication bandwidth has been increasing significantly when messages from a camera, lidar, or radar become a part of safety-relevant functions in modern self-driving cars \cite{hanzalek2023}.

In this paper, we study the decision problem of scheduling non-preemptive jobs on a single machine, where the periods of the jobs are harmonic, i.e., for any two distinct periods among the instance's jobs, it holds that the longer period is divisible by the shorter. We also assume that the schedule has to be strictly periodic: the interval between two occurrences of the same job is fixed and equal to the job period.

$PSPs$ with a harmonic set of periods are easier to solve than the more general case. If the processing times were also harmonic, the problem could be solved in polynomial time, as shown in \cite{korst1996scheduling}. However, our studied problem with harmonic periods and non-harmonic processing times was proven to be strongly NP-complete in \cite{cai1996nonpreemptive}. Schedules with harmonic periods were applied in many industrial settings. This includes FlexRay standard, a communication protocol used in higher-class cars \cite{lukasiewycz2009flexray,dvorak2014multi}. With FlexRay, the goal is to find a periodic schedule for a so-called static segment where the jobs represent the messages and the machine represents the physical communication channel. Furthermore, Boeing's avionics-related problem instances were solved using methods for harmonic periods in \cite{eisenbrand2010solving} as well. 

Due to the NP-hardness of most $PSPs$, authors usually focus on the development of heuristic and approximation algorithms or on using existing exact formalisms and solvers. The $PSPs$ are usually modeled using time-indexed formulations or relative order formulations, implemented using Satisfiability Modulo Theory in \cite{minaeva2022}, or Integer Linear Programming (ILP) in \cite{eisenbrand2010solving}. In \cite{hladik2020complexity}, $PSP$ with harmonic periods with dedicated machines and precedence constraints was tackled using a local search heuristic. The core of their heuristic is a reconstruction function, which schedules each job as early as possible.

Other methods build upon specific characteristics of the $PSP$ with harmonic periods. In this case, the feasible schedule can be visualized using the bin tree representation, shown in \cite{Eisenbrand2010}, or using the 2D packing approach described in this paper. In \cite{Eisenbrand2010}, authors use the compressed bin trees to develop a first fit heuristic to solve a machine minimization problem. Later in \cite{eisenbrand2010solving}, the authors presented an alternative well-performing ILP formulation. Similar first fit heuristic and ILP formulation, used in the context of the FlexRay bus, were shown in \cite{lukasiewycz2009flexray}. The authors focus on minimization of the number of cycles used in the FlexRay communication. In \cite{dvorak2014multi}, a first-fit-based heuristic was utilized to create a FlexRay schedule that could be embedded into multiple models of vehicles. In \cite{marouf2011scheduling}, authors performed a schedulability analysis of both harmonic and non-harmonic $PSPs$ and provided a necessary and sufficient schedulability condition when all the periods are harmonic and unique. 

The crucial correspondence between the $PSP$ with harmonic periods and bin packing problems allows us to use methods developed for the packing and cutting problems. In \cite{fleszar2002onedpinpack}, authors presented a bin-focused heuristic. The proposed solution in each step selects a subset of items that best fits the current bin, packs the items, and continues with another bin. A similar method, generalized for a variable-sized bin packing, was presented in \cite{hauourari2009subsetsumandcolumns}. 

In the 2D case, the guillotine cutting and k-stage packing problems can be considered a generalization of the height-divisible 2D packing later defined in this paper. On the other hand, height-divisible 2D packing is more general than 1D packing. In \cite{ntene2009}, authors provide the overview of guillotine heuristics for strip packing and present their own size alternating stack heuristic, further improved in \cite{ortmann2010levelpack}. The heuristic's main idea is to alter between placing "wide" and "narrow" rectangles to fill space otherwise wasted by simpler heuristics.

With respect to the relevant research, the main contributions of our paper are the following:
\begin{itemize}
    \item We prove the equivalence of the $PSP$ with harmonic periods on a single machine and the height-divisible 2D packing problem, and we show that any solution obtained for the 2D packing problem can be transformed into a canonical form. See Section \ref{sec:correspondence}. 
    \item We present a Constraint Programming (CP) formulation of the problem and compare it with the state-of-the-art ILP model. See Section \ref{sec:exact}.
    \item We present a packing-inspired heuristic, which enables us to solve more highly utilized problem instances in comparison with baseline methods. See Sections \ref{sec:heuristics} and \ref{sec:experiments}. 
\end{itemize}

\section{\uppercase{Problem definition}}
\label{sec:definition}
We consider a set of $n$ independent jobs, $\job{1},\ldots,\job{n}$ to be performed repeatedly on a single machine. Each job $\job{i}$ is characterized by its integer processing time $\proctime{i}$ and its integer period $\pertask{i}$. Periods  belong to a harmonic set $\per{0},\ldots,\per{r-1}$ of size $r$, so that $\forall k\in\{1,\ldots,r-1\}, \per{k}=\base{k}\per{k-1}$ where $\base{k}$ is an integer. We define the least period $\larg=\per{0}$, and we define the hyper-period $\hat\hyperp$ to be the least common multiplier of periods. Due to the harmonic nature of the period set, $\hat\hyperp$ is equal to the longest period $\per{r-1}$.

Any periodic schedule defines for each job $\job{i}$ a start time $first_{i}\le \pertask{i}$ of its first occurrence. The $j$-th occurrence of $\job{i}$ will start  at time  $first_{i}+(j-1)\pertask{i}$, where $ j\ge 1$. We consider here the existence of a feasible schedule. Let $o$ be the starting time of the first job $\job{f}$ with the least period $\larg$. We choose $o$ as the origin. So we define $\start{i}$ such that: $$\start{i}=\left\{\begin{array}{ll}first_{i}-o & if \ first_{i}\ge o\\
 first_{i}+\pertask{i}-o&otherwise\end{array}\right..$$


 We call time interval $[o,o+\hat\hyperp)$ the "observation interval". Notice that for any integer $k\ge 1$, the schedule of time interval $[o+(k-1)\hat\hyperp,o+k\hat\hyperp)$ is the same as the schedule of the observation interval shifted by $(k-1)\hat\hyperp$ time units. In the rest of the paper, we consider without loss of generality that the first occurrence of a job $\job{i}$ starts at $\start{i}$ in the observation interval.
 
Let us now decompose $\start{i}$ with respect to the least period $\larg$:
  $\start{i}= \horis{i}+\vertis{i}\cdot\larg$, with $u_i<\larg$, and integer $\vertis{i}$. Also, $\horis{i}+\proctime{i}\le \larg$, since an occurrence of job $\job{i}$ 
  needs to fit between two occurrences of $\job{f}$ at $o$ and $o+w$.
   
A feasible schedule has to be collision-free. We call collision between two jobs $\job{i},\job{j}$ a situation when two occurrences of these jobs are processed simultaneously.
   
We denote by $\bbase{k}=\prod_{i=1}^{k}\base{i}$, $\bbase{0}=1$, so that:
\begin{equation}
    \per{k}=\bbase{k}\larg.
\end{equation}

We can now state the necessary and sufficient condition for a collision to occur in a periodic schedule.
\begin{lemma}
A periodic schedule induces a collision between two jobs $\job{i},\job{j}$ with periods $\pertask{i},\pertask{j}$ such that $\pernb{i}\le\pernb{j}$ if and only if the two following conditions hold:
\begin{equation}
 \horis{i}<\horis{j}+\proctime{j}\quad and\quad  \horis{j}<\horis{i}+\proctime{i}
    \label{collisionhoriz}
\end{equation}
\begin{equation}
\exists \kappa\in  \mathbb{N}\cap[0,\frac{\bbase{\pernb{j}}}{\bbase{\pernb{i}}}),\quad \vertis{j}=\vertis{i}+\kappa\bbase{\pernb{i}}
    \label{collisionvert}
\end{equation}
\label{lemma:collisionjobs}
\end{lemma}

\begin{proof}
 Suppose that the occurrences $k,k'$ of  $\job{i},\job{j}$ collide. Then  the two following conditions hold:
\begin{multline}
  \start{i}+k\pertask{i}+\proctime{i}> \start{j}+k'\pertask{j}\\ and\ 
     \start{j}+k'\pertask{j}+\proctime{j}>  \start{i}+k\pertask{i}.
\end{multline}

\begin{multline}
\Longleftrightarrow \horis{i}+(\vertis{i}+k\bbase{\pernb{i}})\larg+\proctime{i}>\horis{j}+(\vertis{j}+k'\bbase{\pernb{j}})\larg
\\and\\
\horis{i}+(\vertis{i}+k\bbase{\pernb{i}})\larg<\horis{j}+(\vertis{j}+k'\bbase{\pernb{j}})\larg+\proctime{j}
\end{multline}

\begin{multline}
\Longleftrightarrow \horis{i}+\proctime{i}-\horis{j}>(\vertis{j}-\vertis{i}+k'\bbase{\pernb{j}}-k\bbase{\pernb{i}})\larg
\\
and\\
(\vertis{j}-\vertis{i}+k'\bbase{\pernb{j}}-k\bbase{\pernb{i}})\larg>\horis{i}-\horis{j}-\proctime{j}
\end{multline}

As $\horis{i}+\proctime{i}-\horis{j}<\larg$ and $\vertis{i},\vertis{j}$ are integers, this can occur only if $\vertis{j}-\vertis{i}+k'\bbase{\pernb{j}}-k\bbase{\pernb{i}}\le 0$.

Similarly, as $\horis{i}-\horis{j}-\proctime{j}>-\larg$ the second condition can occur only if $\vertis{j}-\vertis{i}+k'\bbase{\pernb{j}}-k\bbase{\pernb{i}}\ge 0$.

Hence $\vertis{j}-\vertis{i}+k'\bbase{\pernb{j}}-k\bbase{\pernb{i}}= 0$. As $\bbase{\pernb{j}}$ is a multiple of $\bbase{\pernb{i}}$, there exists an integer $\kappa$ such that $\vertis{j}-\vertis{i}=\kappa\bbase{\pernb{i}}$.

Now if this condition holds then a collision occurs if $\horis{j}<\horis{i}+\proctime{i}$ and $\horis{i}<\horis{j}+\proctime{j}$.

\end{proof}

Utilization $\utilization{i}$ of a job $\job{i}$ is the proportion of the time the job uses: $\utilization{i}=\frac{\proctime{i}}{\pertask{i}}$. The utilization of the instance is equal to $\utilinst=\sum_{i=1}^n\utilization{i}$. Obviously, $\utilinst$ is not greater than $1$ in feasible instances.

\section{\uppercase{2D packing and periodic scheduling}}
\label{sec:correspondence}
In this section, we first present a special case of a 2D packing problem and then show the packing problem associated with a periodic scheduling instance. Finally, we prove the equivalence between the two problems by establishing the correspondence between schedule times and 2D coordinates of rectangles.
\subsection{Height-Divisible 2D Packing Problem}
We define the Height-Divisible 2D packing problem. In the rest of the paper, we will use the acronym $HD2D$ packing problem.
An instance of $HD2D$ packing is defined by $n$ rectangles. Each rectangle $\rect{i}$ has width $ \ell_i$ and height $\height{i}$. These rectangles are to be packed in a 2D bin of width $\larg$ and height $\hyperp$. 
We assume that there are $r$  different heights  $H_0\ge \dots\ge H_{r-1}$, each one divides
$\hyperp$ and that they form a harmonic set: each a height $H_k$ divides $H_j$ if $j<k$. Note, that $H_0 = H$ and $H_{r-1} = 1$. 

A feasible packing defines the bottom left corner coordinates $(\horip{i},\vertip{i})$  of the rectangles, so there are no collisions between two rectangles, and they all fit into the bin ($\horip{i}+\ell_i\le w, \vertip{i}+\height{i}\le H$).

A packing is said to be {\bf height divisible} if the height coordinate $\vertip{i}$ is divisible by the height of the rectangle:
$$\vertip{i}\mod \height{i}=0.$$

So, the $HD2D$ packing problem is to find a feasible height-divisible packing. We can now state the conditions of a collision between two rectangles in a height-divisible packing.

\begin{lemma} In a height-divisible packing, a collision  between rectangles $\rect{i}$ and $\rect{j}$  such that $\height{j}\le\height{i}$ occurs iff the two following conditions hold:
 \begin{equation}\
  \horip{j}<\horip{i}+\ell_{i},\quad and\quad \horip{i}<\horip{j}+\ell_{j}
  \label{eq;collisionhorizpacking}
\end{equation}
\begin{equation}
 \vertip{i}\le \vertip{j}<\vertip{i}+\height{i}, 
   \label{eq:collisionpacking}
 \end{equation}
\label{lemma:collisionpacking}
\end{lemma}
\begin{proof}
Obviously, a collision occurs if: 
\begin{align}
 \label{vertipcollision}  
 \horip{i}-\ell_{j}<\horip{j}<\horip{i}+\ell_{i} \\ 
 \vertip{i}-\height{j}<\vertip{j}<\vertip{i}+\height{i}
\label{horipcollision}
\end{align}

Assuming $\height{j}\le\height{i}$,  $\height{j}$ divides $\height{i}$ and also divides $\vertip{j}$ and $\vertip{i}$. Consequently, the inequality $\vertip{i}-\height{j}<\vertip{j}$ reformulates to $\vertip{i}\leq\vertip{j}$ and the other inequalities remain the same.
\end{proof}

\subsection{$HD2D$ Packing Instance Associated with a $PSP$ Instance}\label{sec:equivalence}
We can now define a 2D packing instance associated with a periodic scheduling instance with harmonic periods.
Consider an instance $I$ of the scheduling problem. We define an associated instance $\phi(I)$ of the $HD2D$ packing problem.

We set  $\hyperp=\frac{\hat\hyperp}{\larg}=\bbase{r-1}$. To each job $\job{i}$ 
we associate a rectangle $\rect{i}$ of width $\ell_i=\proctime{i}$ and height $\height{i}=\frac{\hat\hyperp}{\pertask{i}}$, to fit in a bin of width $\larg$ and height $\hyperp$.

\begin{figure}
\centering
   \begin{subfigure}[b]{0.42\textwidth}
   \includegraphics[width=0.99\textwidth]{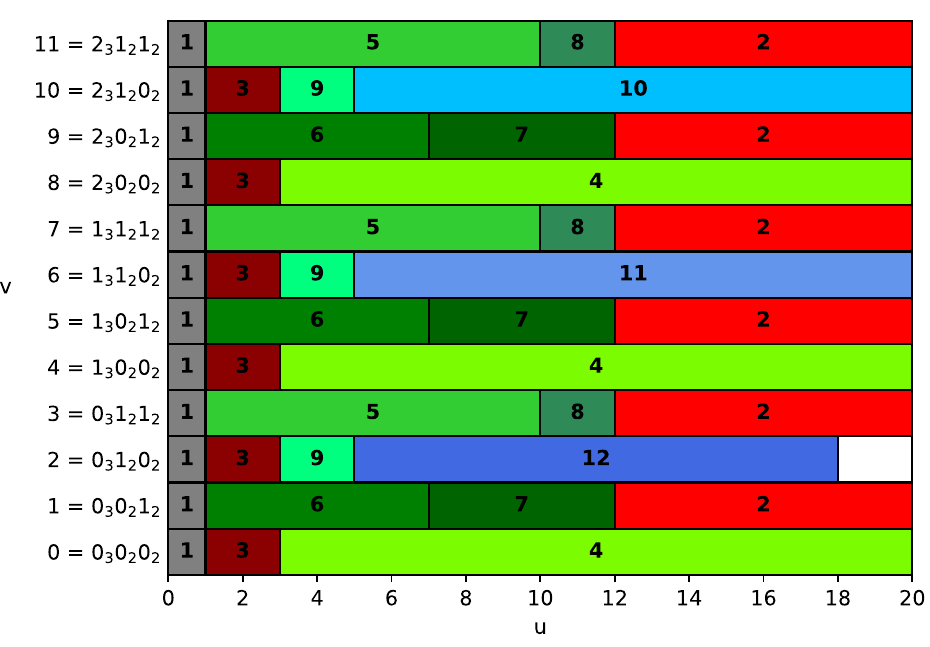}
   \caption{Periodic scheduling problem}
   \label{fig:sample_periods} 
\end{subfigure}
\vskip\baselineskip
\begin{subfigure}[b]{0.42\textwidth}
   \includegraphics[width=0.99\textwidth]{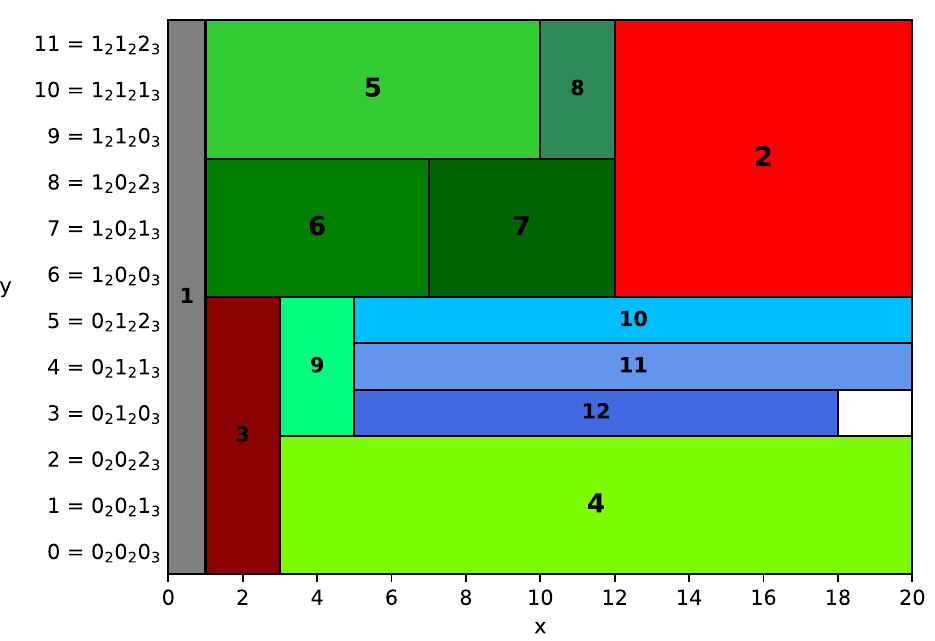}
   \caption{Height-divisible 2D packing problem}
   \label{fig:sample_placement}
\end{subfigure}

\caption{Solutions to $PSP$ and $HD2D$ packing problems, with corresponding jobs and rectangles presented using the same colors. Row indices use the mixed-radix system.}
\label{fig:sample_figures}
\end{figure}

To give an intuition of the correspondence between a periodic schedule and a height-divisible packing, we first develop it on an example in Figures \ref{fig:sample_figures}. The main idea is to arrange the periodic schedule in rows, as shown in Figure~\ref{fig:sample_periods}. A row indexed by $v$ represents the schedule of the time window $[vw,(v+1)w)$ of the observation interval.

The instance contains 12 jobs $\job{1},\dots,\job{12}$. $\job{1}$ is the only job with the shortest period $\pertask{1} = \per{0} = 20$. Jobs $\job{2}, \job{3}$ have period equal to 40, jobs $\job{4},\dots,\job{9}$ have period equal to 80, and finally, jobs $\job{10},\job{11},\job{12}$ have period equal to 240. The chart of the periodic schedule therefore has width $w=\per{0} = 20$, and its height is equal to $\hat\hyperp / w = \bbase{3} = 12$. The 2D bin of the corresponding $HD2D$ packing problem shown in Figure~\ref{fig:sample_placement} has the same width $w$ and height $H$.

Figure~\ref{fig:sample_periods} shows the chart of the feasible periodic schedule over a single observation interval of length $\per{3} = 240$. Origin $0$ is set by the start time of job $\job{1}$, and we stack each time window of length $w=20$ on top of the previous one (see the rows in Figure~\ref{fig:sample_periods} with their three-digit row numbers using mixed-radix system formally defined in Section~\ref{sec:ekvi_proof}). We can see that, e.g., job $\job{4}$ has a period of 80, as its subsequent occurrence is at the fourth row from its previous occurrence. Note that jobs with the same period use similar colors. 

Figure~\ref{fig:sample_placement} shows the corresponding height-divisible packing view of the same solution. 
The packing is obtained by interchanging the rows as follows:
we swap the three-digit row numbers in  Figure~\ref{fig:sample_periods} such that the most significant digit becomes the least significant digit and vice versa. Finally, we obtain Figure~\ref{fig:sample_placement} by ordering the rows according to their new three-digit numbers.
For example, the row with $\job{11}$ has number $1_3 1_2 0_2 = 6$ in Figure~\ref{fig:sample_periods} and after the swap, the same row has number $0_2 1_2 1_3 = 4$ in Figure~\ref{fig:sample_placement}.

We can see the occurrences of the original jobs have been merged, and the dimensions of the resulting rectangles correspond to those described in Section~\ref{sec:equivalence}. But this is just an observation based on one example, which does not guarantee a compact shape of rectangles in $HD2D$ packing. The formal proof of the equivalence of both problems is derived in Section~\ref{sec:ekvi_proof} and uses positions of jobs and rectangles rather than three-digit numbers of rows used in this Section~\ref{sec:equivalence}.

Note that there exists a bin tree representation of $PSP$ with harmonic periods. It was used in \cite{Eisenbrand2010} and is closely related to our $HD2D$ packing perspective. The bin tree represents the $PSP$ with a tree, with each level of nodes corresponding to one of the periods. Examples of the bin tree and $HD2D$ packing representations are shown in Figures 7, 8 in \cite{minaeva2022}.

\subsection{Flip Operation and Equivalence between $HD2D$ Packing Problem and $PSP$}\label{sec:ekvi_proof}
In this section, we will formally introduce a flip operation that allows us to get a bijection between feasible periodic schedules and height-divisible packing of the associated packing instance.

\begin{theorem}\label{theor:flip}
To any periodic feasible schedule of the instance $I$, a feasible 2D height-divisible packing of $\phi(I)$ is associated.
\begin{equation}
\forall \job{i},\quad \left\{\begin{array}{ll}\horip{i}&=\horis{i}\\
\vertip{i}&=\height{i}\cdot\flip{(\vertis{i},\pernb{i},{\bf\base{}})}\end{array}\right.
\label{sched2pack}
\end{equation}
Conversely, from a 2D height-divisible packing  of $\phi(I)$ we can define a feasible schedule  as follows:
\begin{equation}
\forall \job{i},\quad \left\{\begin{array}{ll}\horis{i}&=\horip{i}\\
\vertis{i}&=\flip{(\frac{\vertip{i}}{\height{i}},\pernb{i},\bflip{({\bf\base{}},\pernb{i})})}\end{array}\right.
\label{pack2sched}
\end{equation}
\end{theorem}

We first introduce the theoretical framework used to prove Theorem~\ref{theor:flip}.  Let $y$ be any natural number.  It is known that $y< H$ can be decomposed uniquely according to the mixed radix numerical system ${\bf\base{}}=(\base{1},\ldots,\base{r-1})$ as follows:
$$y=y_1\bbase{0}+y_2\bbase{1}+ y_3\bbase{2}+\ldots+y_{r-1}\bbase{r-2}$$
with $\forall i> 0, y_i<\base{i}$. This decomposition is usually denoted with the most significant digits to the left as follows: \begin{equation}
    [y]_{{\bf\base{}}}=[y_{r-1}]_{\base{r-1}}\ldots[y_1]_{\base{1}}
\end{equation}

Notice that in the usual base decomposition (for example, base 2 for a binary decomposition), the components of the base vector ${\bf\base{} }$ are all equal (for example, equal to 2 for the binary decomposition).


Let us generalize the transformation proposed by \cite{lukasiewycz2009flexray}
for the binary decomposition of a number by defining two operators:
Let $\bflip{({\bf\base{}},k)}$ be the operator that flips the $k$ first component of the base vector $\base{}$:
\begin{equation}
    \bflip{({\bf\base{}},k)}=(\base{k},\base{k-1},\ldots,\base{1},\base{k+1}\ldots,\base{r-1})\label{def:bflip}
\end{equation}

Let now $\flip(y,k,{\bf\base{}})$ be the number constructed by flipping the $k$ least significant digits of the decomposition $[y]_{{\bf\base{}}}$ to give a number expressed with respect to base vector ${\bf\base{}'}=\bflip{({\bf\base{}},k)}$:
\begin{equation}
[\flip(y,k,{\bf\base{}})]_{{\bf\base{}'}}=[y_{r-1}]_{\base{r-1}}\ldots[y_{k+1}]_{\base{k+1}} [y_1]_{\base{1}}\ldots[y_k]_{\base{k}}
    \label{def:flip}
\end{equation}
Notice that the $r-1-k$ most significant digits are not modified.

Hence, we have:
\begin{multline}
    \flip(y,k,{\bf\base{}})=y_k\cdot 1+y_{k-1}\cdot\base{k}+\ldots+y_1(\base{2}\cdots\base{k})+\\ y_{k+1}(\base{1}\cdots\base{k})+\ldots+y_r(\base{1}\cdots\base{r-2})\end{multline}

	Example: consider  the base vector ${\bf\base{}}=(2,2,3)$ used in Figure~\ref{fig:sample_periods}. So $H=12$. Let us consider $y=10$.  Its decomposition along ${\bf\base{}}$ is $[y]_ {{\bf\base{}}}=[2]_{3}[1]_{2}[0]_{2}=0+1\cdot 2+2\cdot 4$. $\bflip({\bf\base{}},3)=(3,2,2)$. $\flip(y,3,{\bf\base{}})=[0]_{2}[1]_{2}[2]_{3}=2+1\cdot 3+0\cdot 6=5$.

This flip operation has some important properties that will be used to transform the scheduling problem into an equivalent packing problem.  The following lemma expresses the reversibility of the flip and the fact that equidistant integers (with distance $\bbase{k-1}$) become consecutive integers after the flip operation.

\begin{lemma}
For any integer $y<H$, \begin{equation}\flip{(\flip{(y,k,{\bf\base{}})},k,\bflip{({\bf\base{}},k)})}=y\label{eq:doubleflip}\end{equation}
 if $[y_k]_{\base{k}}<\base{k}-1$, then \begin{equation}\flip{(y+\bbase{k-1},k,{\bf\base{}})}=\flip{(y,k,{\bf\base{}})}+1\label{eq:sommebk}\end{equation}
 and if $[y_k]_{\base{k}}>0$,
 \begin{equation}\flip{(y-\bbase{k-1},k,{\bf\base{}})}=\flip{(y,k,{\bf\base{}})}-1\label{eq:diffbk}\end{equation}
 If $y$ is a multiple of $\bbase{k}$ ,  $z<\bbase{k}$ and  $k'\ge k$ then
 \begin{equation}
 \flip(y+z,k',{\bf\base{}})=\flip(y,k',{\bf\base{}})+\flip(z,k',{\bf\base{}})\label{eq:sommeflip}
 \end{equation}
\label{lemma:flip}
\end{lemma}
\begin{proof}
Obviously, $\bflip(\bflip({\bf\base{}},k),k)={\bf\base{}}$, since a double flip of the $k$ first digits lead to the initial digits. This proves \eqref{eq:doubleflip}.  Let us consider \eqref{eq:sommebk}. If we add $\bbase{k-1}$ to $y$, the digit $[y_k]_{\base{k}}$ increases by one unit, so after the flip operation, this digit will be the first; hence, the corresponding number increases by $1$.  Similar arguments hold for \eqref{eq:diffbk}.  Finally, if $y$ is a multiple of $\bbase{k}$, its $k$ least significant digits are null with respect to base ${\bf\base{}}$. If $z<\bbase{k}$  its digits from $k+1$  to the last one are null, so the sum $y+z$ consists in the $k$ first digits of $z$ followed by the next digits of $y$. 
If $k'\ge k~$ $\flip(y,k',{\bf\base{}})$ has $k'-k$ first flipped digits from the digits $k$ to $k'$ of $y$ and  $\flip(z,k',{\bf\base{}})$  starts with $k'-k$ null digits followed by the flipped $k$ digits of $z$. So, the sum is exactly the flipped digits from $y+z$.
\end{proof}

We are now able to prove the equivalence between $PSP$ and $HD2D$ packing problem.

\paragraph{Proof of Theorem~\ref{theor:flip}}
\begin{proof}
The proof relies on the properties of the flip operation stated in Lemma~\ref{lemma:flip} and the expressions of collisions stated in Lemma~\ref{lemma:collisionjobs} and \ref{lemma:collisionpacking}.

Assume that there is a feasible schedule of instance $I$.
Now, assume that a collision occurs in the associated 2D packing between rectangles $\rect{i}$ and $\rect{j}$ such that $\height{i}\ge\height{j}$,
so that $\pertask{i}\le \pertask{j}$. 

So according to equation~\eqref{eq;collisionhorizpacking}, and as $\horip{i}=\horis{i},\horip{j}=\horis{j}$ we have $\horis{j}<\horis{i}+\proctime{i}$ and $\horis{i}<\horis{j}+\proctime{j}$. This implies that the  condition~\eqref{collisionhoriz} of Lemma~\ref{lemma:collisionjobs} holds, and thus, as there is no collision in the schedule, the second condition cannot hold: $\vertis{j}-\vertis{i}$ cannot be a multiple of $\bbase{\pernb{i}}$.

Now according to Lemma \ref{lemma:collisionpacking}, equation~\eqref{vertipcollision}, the collision in the packing implies that:
$$\vertip{i} \le \vertip{j}<\vertip{i}+\height{i}.
$$

Assume that $ \vertip{j}=\vertip{i}+\Delta$. Notice that as $\height{j}$ divides $\height{i}$, and as $\height{j}$ divides $ \vertip{j}$, it should also divide $\Delta$. Moreover:
\begin{align*}
   \vertip{j}= \height{j}\cdot\flip{(\vertis{j},\pernb{j},{\bf\base{}})}=   \height{i}\cdot\flip{(\vertis{i},\pernb{i},{\bf\base{}})}+\Delta\\
    \flip{(\vertis{j},\pernb{j},{\bf\base{}})}=\frac{\height{i}}{\height{j}}\flip{(\vertis{i},\pernb{i},{\bf\base{}})}+\frac{\Delta}{\height{j}}\\
    with\ \frac{\Delta}{\height{j}}<\frac{\height{i}}{\height{j}}
\end{align*}

Now according to Lemma~\ref{lemma:flip}, 
\begin{align*}
\vertis{j}=\flip{(\frac{\vertip{j}}{\height{j}},\pernb{j},\bflip{({\bf\base{}},\pernb{j})})}\\
=\flip{(\frac{\height{i}}{\height{j}}\flip{(\vertis{i},\pernb{i},{\bf\base{}})}+\frac{\Delta}{\height{j}},\pernb{j},\bflip{({\bf\base{}},\pernb{j}}))}
\end{align*}

Observe that $\frac{\height{i}}{\height{j}}=\frac{\pertask{j}}{\pertask{i}}=\base{\pernb{i}+1}\ldots\base{\pernb{j}}$.  
And $\frac{\Delta}{\height{j}} <\frac{\height{i}}{\height{j}}=\base{\pernb{i}+1}\ldots\base{\pernb{j}}$. 
So applying Lemma~\ref{lemma:flip} with base $\bflip{({\bf\base{}},\pernb{j})}$ we can say that:

\begin{align*}
\vertis{j}= 
\flip{(\frac{\height{i}}{\height{j}}
\flip{(\vertis{i},\pernb{i},{\bf\base{}})}
,\pernb{j},\bflip{({\bf\base{}},\pernb{j})})}\\
+\flip{(\frac{\Delta}{\height{j}},\pernb{j},\bflip{({\bf\base{}},\pernb{j})})}
\end{align*}

In the base representation $\bflip{({\bf\base{}},\pernb{j})}$ the  
$\pernb{j}-\pernb{i}$ least significant digits of $\frac{\height{i}}{\height{j}}\flip{(\vertis{i},\pernb{i},{\bf\base{}})}$ equal $0$. And the next $\pernb{i}$ digits are the ones of $\vertis{i}$ flipped. So that 
the  $\pernb{i}$ least significant digits of $\flip{(\frac{\height{i}}{\height{j}}\flip{(\vertis{i},\pernb{i},{\bf\base{}})},\pernb{j},\bflip{({\bf\base{}},\pernb{j}}))}$ in the base representation ${\bf\base{}}$ are equal to the  $\pernb{i}$ least significant digits of $\vertis{i}$ in base representation ${\bf\base{}}$ followed by digits $0$. Hence 
$\vertis{i}=\flip{(\frac{\height{i}}{\height{j}}\flip{(\vertis{i},\pernb{i},{\bf\base{}})},\pernb{j},\bflip{({\bf\base{}},\pernb{j}}))}$.

 Moreover the number $\flip{(\frac{\Delta}{\height{j}},\pernb{j},\bflip{({\bf\base{}},\pernb{j}}))}$ has its  $\pernb{i}$ least significant digits equal to $0$ in the base representation ${\bf\base{}}$. So that it is a multiple of $\bbase{\pernb{i}}$: 
\begin{align}
    \vertis{j}= \vertis{i}+ \kappa \bbase{\pernb{i}}.
\end{align}

 Since we know that $\vertis{j}<\frac{\pertask{j}}{\larg}=\bbase{\pernb{j}}$, we have $\kappa<\frac{\bbase{\pernb{j}}}{\bbase{\pernb{i}}}$.  So condition \eqref{collisionvert} of Lemma~\ref{lemma:collisionjobs} holds, which contradicts our assumption that no collision occurs.
 
 Conversely, assume that we have a feasible packing and let us assume that there is a collision in the associated schedule between jobs $\job{i}$ and $\job{j}$ with $\pertask{i}\le \pertask{j}$.  The first condition of Lemma~\ref{lemma:collisionjobs} holds so that the first condition of Lemma~\ref{lemma:collisionpacking} holds either.  This implies that the  condition \eqref{eq:collisionpacking} of Lemma~\ref{lemma:collisionpacking} does not hold, and either $\vertip{i}>\vertip{j}$ or $\vertip{j}\ge \vertip{i}+\height{i}$.
 
 But if a collision occurs in the schedule, we should have:
 $\vertis{j}=\vertis{i}+\kappa\bbase{\pernb{i}}$.


So,
\begin{align*}
\frac{\vertip{j}}{\height{j}}=\flip{(\vertis{j},\pernb{j},{\bf\base{}})}\\
=\flip{(\vertis{i}+ \kappa \bbase{\pernb{i}},\pernb{j},{\bf\base{}})}\\
=\flip{(\flip{(\frac{\vertip{i}}{\height{i}},\pernb{i},\bflip{({\bf\base{}},\pernb{i})})}+ \kappa \bbase{\pernb{i}},\pernb{j},{\bf\base{}})}\end{align*}

Notice that as  $\vertip{i}< H$ so that $\frac{\vertip{i}}{\height{i}}<\bbase{\pernb{i}}$ and thus  when performing the operation $\flip{(\frac{\vertip{i}}{\height{i}},\pernb{i},\bflip{({\bf\base{}},\pernb{i})})}$, we get a number still less than $\bbase{\pernb{i}}$. We can then apply Lemma~\ref{lemma:flip}, and we get: 

\begin{align*}
    \flip{(\flip{(\frac{\vertip{i}}{\height{i}},\pernb{i},\bflip{({\bf\base{}},\pernb{i})})}+ \kappa \bbase{\pernb{i}},\pernb{j},{\bf\base{}})}\\=\flip{(\flip{(\frac{\vertip{i}}{\height{i}},\pernb{i},\bflip{({\bf\base{}},\pernb{i})})},\pernb{j},{\bf\base{}})} + \kappa\\
    =\frac{\bbase{\pernb{j}}}{\bbase{\pernb{i}}}\cdot\frac{\vertip{i}}{\height{i}}+\kappa
    \end{align*}
 
Now $\height{i}\bbase{\pernb{i}}=\height{j}\bbase{\pernb{j}}=H$, so that: 
\begin{equation}
    \vertip{j}=\vertip{i}+\kappa\cdot\height{j}
\end{equation}

Now we know that $\kappa< \frac{\bbase{\pernb{j}}}{\bbase{\pernb{i}}}=\frac{\height{i}}{\height{j}}$. So $\vertip{j}<\vertip{i}+\height{i}$ and condition \eqref{eq:collisionpacking}  of Lemma~\ref{lemma:collisionpacking} holds, a contradiction.
 
\end{proof}


\subsection{Canonical Form of a $HD2D$ Packing}\label{sec:canonical}
In this section, we prove that any $HD2D$ packing can be put in a canonical form for which the packed rectangles' heights decrease from the left to the right. 




\begin{definition}
A height-divisible packing is said to be in canonical form if for all $i,j\in\{1,\ldots,n\}$
\begin{equation}\height{i}>\height{j},\quad \vertip{j}\ge\vertip{i},\quad \vertip{j}+\height{j}\le \vertip{i}+\height{i}\Longrightarrow \horip{i}+\ell_i\le\horip{j}
 \label{eq:canonical}
\end{equation}
\end{definition}

Such a packing corresponds to a schedule of the observation interval where in each interval $[k\larg,(k+1)\larg)$ (such interval corresponds to the $k$-th row in Figure~\ref{fig:sample_periods}), jobs are scheduled in increasing order of periods.
\begin{theorem}
If there exists a feasible packing of width $\ell$, then there exists a feasible canonical packing of width $\ell$.
\end{theorem}
\begin{proof}
By induction on  $r$ the number of different heights of rectangles. 
Of course, if $r=1$, the theorem holds.  Assume this is true for any height-divisible packing with at most $r$ different heights.  Consider an instance with $r+1$ different heights.  We claim that the rectangles of maximal height $\hyperp = H_0$ 
can be packed at the left of the bin, and the other rectangles can be right-shifted accordingly.  Let $\ell'$ be the sum of the lengths of the rectangles of height $H_0$. 
Consider $H_1$ 
the second largest height. 
If $\rect{j}$ is a rectangle of height $\height{j}\le H_1$, we know that its vertical coordinate is a multiple of $\height{j}$.  So  there exists a non-negative integer $q$ such that $\vertip{j}=q \height{j} = q'H_1+  \vertip{j}'$, with $\vertip{j}'\ge 0$ and $\vertip{j}'+\height{j}\le H_1 $. 

So we can consider $\frac{H_0}{H_1}$ 
so-called sub-bins of height $\hyperp_1$  and width $\ell-\ell'$ in which all remaining rectangles are packed.  Here, rectangle $\rect{j}$ is packed in sub-bin $q'$.  The set of rectangles packed into a sub-bin has at most $r$ different heights.  By induction, all the rectangles packed in a sub-bin can be packed with a canonical packing.
\end{proof}

\begin{figure}
\centering
   \includegraphics[width=0.42\textwidth]{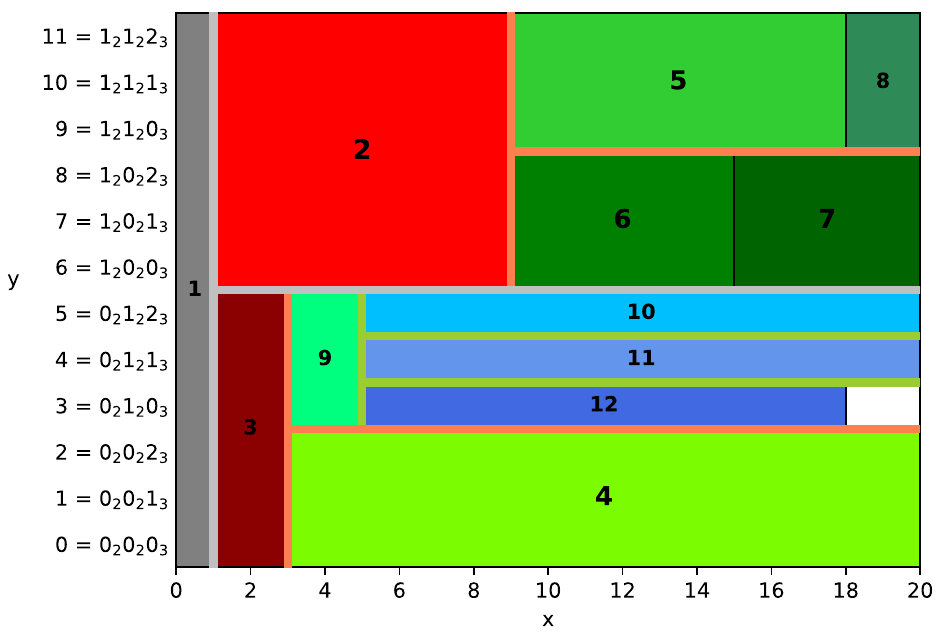}
   \caption{Canonical form of the packing shown in Figure~\ref{fig:sample_placement}. Sub-bin division is highlighted by T-like contours.}
   \label{fig:sample_placement_canonical} 
\end{figure}

The canonical form of the packing shown in Figure~\ref{fig:sample_placement} is presented in Figure~\ref{fig:sample_placement_canonical}.  The canonical form also shows sub-bins crucial for heuristics' description.  The original 2D bin of size $w \times \hyperp$ starts from the left as a single sub-bin of height $H$.  Once all rectangles with height $H_0 = H$ are packed (grey rectangle $R_1$ in Figure~\ref{fig:sample_placement_canonical}), the remaining free space to the right of the parent sub-bin is divided to $\frac{H_0}{H_1}$ sub-bins of height $H_1$.  Once rectangles with height $H_1$ are packed (reddish rectangles $R_2,R_3)$, we divide each sub-bin into $\frac{H_1}{H_2}$ sub-bins of height $H_2$. This continues until sub-bins of height $H_{r-1} = 1$ are created, and the rectangles with the smallest height are packed. 

In Figure~\ref{fig:sample_placement_canonical}, the lower sub-bin of height 6 contains rectangles $R_3, R_4, R_9, R_{10}, R_{11}, R_{12}$. The division of sub-bins is also highlighted by the colored T-like contours.  For example, the sub-bin of height 6, where rectangle $R_2$ is packed, divides itself into two sub-bins, each containing two rectangles of height 3.  Notice that the number of all potential sub-bins of height $H_k$ is equal to $\frac{H_0}{H_k}$, but there may be sub-bins unoccupied by any rectangle of such height. 

\section{\uppercase{CP model}}\label{sec:cp}
\label{sec:exact}
We use the bin formulation, found in \cite{lukasiewycz2009flexray,eisenbrand2010solving}, to model and solve the single machine $PSP$ with harmonic periods. We formulate the problem from the $HD2D$ packing perspective, and the solution to the $PSP$ is obtained using the equations from Section~\ref{sec:ekvi_proof}. The bin formulation only assigns the rectangles to suitable sub-bins, as described in the previous section. Once such assignment is obtained, the horizontal coordinates of rectangles are determined by processing them ordered non-increasingly by their height, packing them as far left as possible in their assigned sub-bin, thus creating a solution in canonical form.

We model the bin formulation using CP. The reasons are the recent advancements of the CP solvers, especially in the scheduling domain, the ease of the modeling, and powerful global constraints. We rely on the $\mathbf{pack}$ constraint \cite{Shaw2004cpconstraint}. The constraint $\mathbf{pack}(\mathbf{L}, \mathbf{a}, \mathbf{l})$ couples together list $\mathbf{L}$ of groups' load variables, list $\mathbf{a}$ of items' assignment variables, and list $\mathbf{l}$ of items' sizes, so: 
\begin{align}
    &a_i \in \lbrace 0,\dots,|\mathbf{L}| - 1\rbrace& \forall i \in \lbrace 0,\dots, |\mathbf{a}| - 1\rbrace\\
    & L_i = \sum_{\forall j\,:\, a_j = i} l_j & \forall i \in \lbrace 0,\dots,|\mathbf{L}| - 1\rbrace
\end{align}

Thus, each item is assigned to one of the groups, and each group's load variable contains the sum of its assigned items' sizes. We can impose additional constraints to control the maximum loads.

To model the $HD2D$ packing problem, we use one pack constraint for each distinct height $H_k$. This pack constraint uses load variables $\mathbf{L}^k =  (L_0^k,\dots,L^k_{H_0/H_k - 1})$  to model all available sub-bins of height $H_k$. The sub-bins are indexed from bottom to top. Lists of widths $\mathbf{l}^k$ and assignment variables $\mathbf{a}^k$ are associated with all rectangles with height $h_j = H_k$. With these definitions, the used pack constraints are:
\begin{align}
	& \mathbf{pack}(\mathbf{L}^k, \mathbf{a}^k, \mathbf{l}^k) & \forall k \in \{0,\dots,r-1\} 
\end{align}

Such constraint distributes the rectangles with height $H_k$ among sub-bins with height $H_k$, and the load variable $L^k_i$ is equal to the sum of widths of rectangles that are assigned to $i$-th sub-bin of height $H_k$. Then, for each one of the $H_0$ rows in the packing, we sum together the loads of sub-bins that intersect such row (see Figure~\ref{fig:sample_placement_canonical}). We ensure that such sum does not exceed the width of the 2D bin $w$:
\begin{align}
     &\sum_{k=0}^{r-1} L^k_{\lfloor i / H_k \rfloor} \le w & \forall i \in \{0,\dots H_0 -1\}  
\end{align}

These two sets of constraints are sufficient to model the problem. Once solved, the assignment of rectangles to sub-bins is determined from values of variables $\mathbf{a}^k$.

\section{\uppercase{Heuristics}}\label{sec:heur_start}
\label{sec:heuristics}
While the exact approaches completely search the solution space, their performance diminishes with an increasing number of jobs to schedule. On the other hand, even simple heuristics are often able to find a solution for large instances \cite{hladik2020complexity}. 

All studied heuristics process the jobs one by one, ordered first non-decreasingly by their period and then non-increasingly by their processing times. 
The period ordering (known as rate-monotonic in the time-triggered community) is natural, several heuristics rely on it, and it produces a solution in canonical form. The secondary ordering by processing times is used by the bin packing and scheduling approximation algorithms (e.g., in \cite{lpt}).

\subsection{Baseline Heuristics}\label{sec:baseline}
\subsubsection{Time-wise First Fit}\label{sec:time}
In \cite{hladik2020complexity}, authors originally tackled the problem with dedicated machines and precedence constraints. Their algorithm schedules each job as soon as possible, minimizing the job's start time in the yet-unfinished schedule. The method does not rely on the rate-monotonic order of jobs, but the outlined order was used as an initial solution of their simulated annealing scheme. Due to the time-oriented aspect of the heuristic, we denote it \textbf{T-FF}.

\subsubsection{Spatial First Fit and Best Fit}\label{sec:spatial}
In \cite{Eisenbrand2010}, the first fit is also used. However, the rectangles of the corresponding $HD2D$ packing problem are packed to the lowest available sub-bin instead. We refer to this approach as a spatial first fit. Furthermore, we also tested best fit as a sub-bin selection policy; thus, we obtained two distinct variants of the algorithm, denoted \textbf{S-FF} and \textbf{S-BF} respectively. 

\subsubsection{Longest Processing Time First}\label{sec:lpt}
The last baseline approach is the well-known Longest Processing Time First list scheduling algorithm for parallel machines scheduling \cite{lpt}. We follow the same spatial approach of Section~\ref{sec:spatial}, but for the current job, its corresponding rectangle is packed into the least occupied sub-bin. We denote this method \textbf{LPT}.

\subsection{Rectangle-Guided First Fit}
We observed that the baseline heuristics often cannot schedule infrequent long jobs, whose corresponding rectangles are wide and low. These rectangles are packed later in the process when the packing is mostly finished. Following this observation, we designed our heuristic, called rectangle-guided first fit, using the $HD2D$ packing perspective. The algorithm consists of two phases. In the first phase, so-called dummy rectangles are created, and in the second phase, the rectangles (including dummy rectangles) are packed using a modified first fit heuristic.

Partly inspired by "wide" and "narrow" rectangles of \cite{ntene2009,ortmann2010levelpack}, the dummy rectangles are created for each distinct height $H_k$ of the problem. For $H_{r-1} = 1$, there are no dummy rectangles. For each subsequent height $H_k$, the dummy rectangles are constructed based on both real and dummy rectangles of height $H_{k+1}$. As the newly created rectangles are taller, they will be processed earlier by the heuristic and will reserve space for the smaller ones. However, their only purpose is to provide a look-ahead; once all the rectangles with height $H_k$ are packed, we remove all the dummy rectangles with height $H_k$ from the sub-bins and continue with packing the rectangles of height $H_{k+1}$.

\subsubsection{On Construction of Dummy Rectangles}
Dummy rectangle of height $H_k$ and width $l$ essentially consists of $H_k/H_{k+1}$ bins of height $H_{k+1}$, each with width $l$. Our aim is to create dummy rectangles by filling these bins with the smaller rectangles of height $H_{k+1}$, so the new rectangles roughly capture how the smaller rectangles could be packed. We create the rectangles heuristically using 1D bin packing. To avoid confusion with the 2D packing, we call the one-dimensional bins {\bf bags}. We propose two approaches for constructing dummy rectangles.

\begin{figure}
\centering
\begin{subfigure}[b]{0.23\textwidth}
   \includegraphics[width=0.999\textwidth]{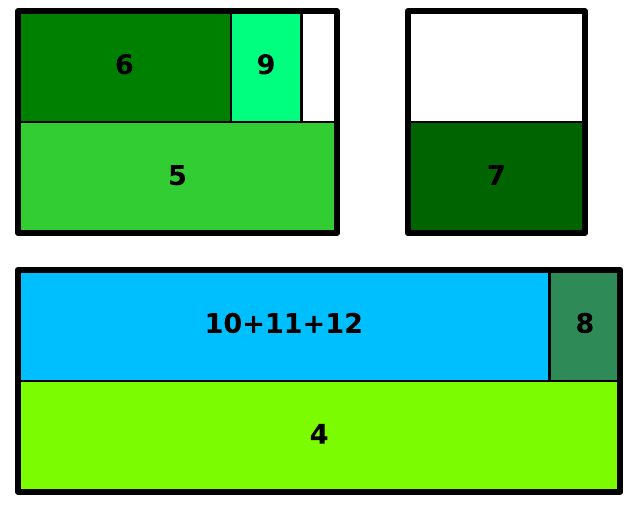}
   \caption{Pessimistic construction}
   \label{fig:pess_rectangles}
\end{subfigure}
\vskip\baselineskip
   \begin{subfigure}[b]{0.23\textwidth}
   \includegraphics[width=0.999\textwidth]{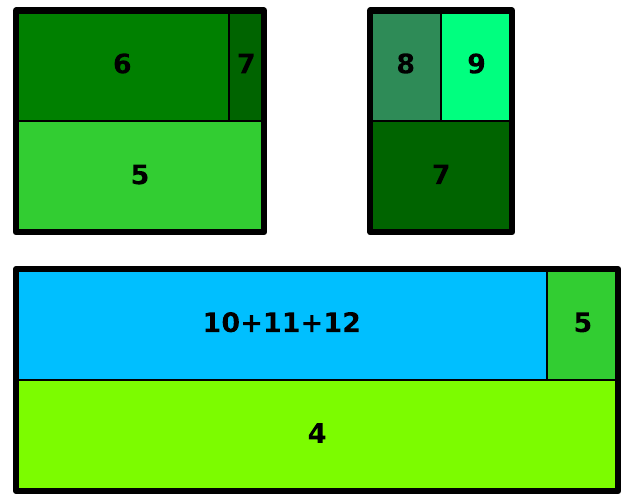}
   \caption{Optimistic construction}
   \label{fig:opt_rectangles} 
\end{subfigure}
\caption{Bag view of dummy rectangles for height $H_1=6$ constructed for instance from Section~\ref{sec:equivalence}.}
\label{fig:constructed_rectangles}
\end{figure}

The pessimistic approach sorts the rectangles with height $H_{k+1}$ by their width. We start with an empty list of bags. If a rectangle $R_i$ with width $l_i$ fits in one of the existing bags, it is packed according to the best fit policy. If not, a new dummy rectangle of height $H_{k}$ is created, with width $l_i$. Simultaneously, $H_k/H_{k+1}$ bags of size $l_i$  are created, and $R_i$ fully fills one of them. The process repeats until all the rectangles are packed in the bags. Since many bags may contain holes, not all constructed dummy rectangles are completely filled with the smaller ones. This is shown in Figure~\ref{fig:pess_rectangles}. Three dummy rectangles of height 6 were created from rectangles of height 3, but two bags associated with two narrower rectangles were not fully filled. Note that the blue rectangle is the dummy rectangle created for $H_2=3$. We refer to our algorithm using pessimistic construction as \textbf{RG-FF-PES}.

\begin{figure*}
\centering
   \begin{subfigure}[b]{0.3\textwidth}
   \includegraphics[width=0.99\textwidth]{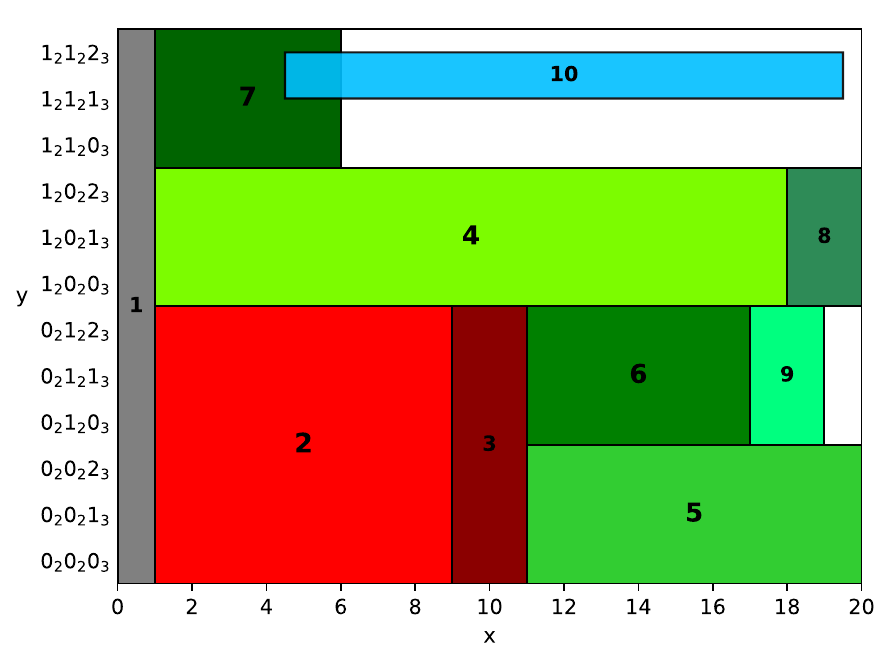}
   \caption{T-FF}
   \label{fig:tff} 
\end{subfigure}
\begin{subfigure}[b]{0.3\textwidth}
   \includegraphics[width=0.99\textwidth]{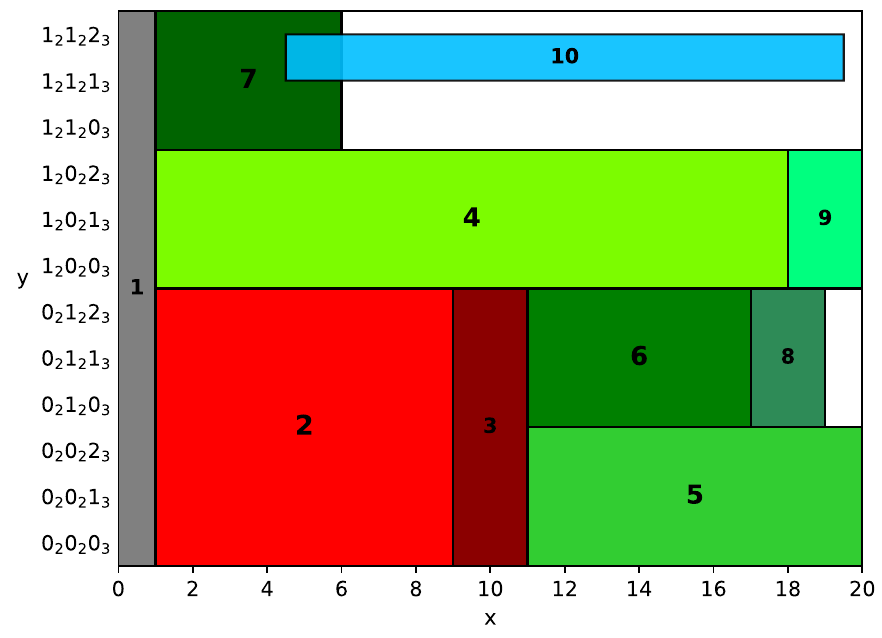}
   \caption{S-FF}
   \label{fig:sbf}
\end{subfigure}
\begin{subfigure}[b]{0.3\textwidth}
   \includegraphics[width=0.99\textwidth]{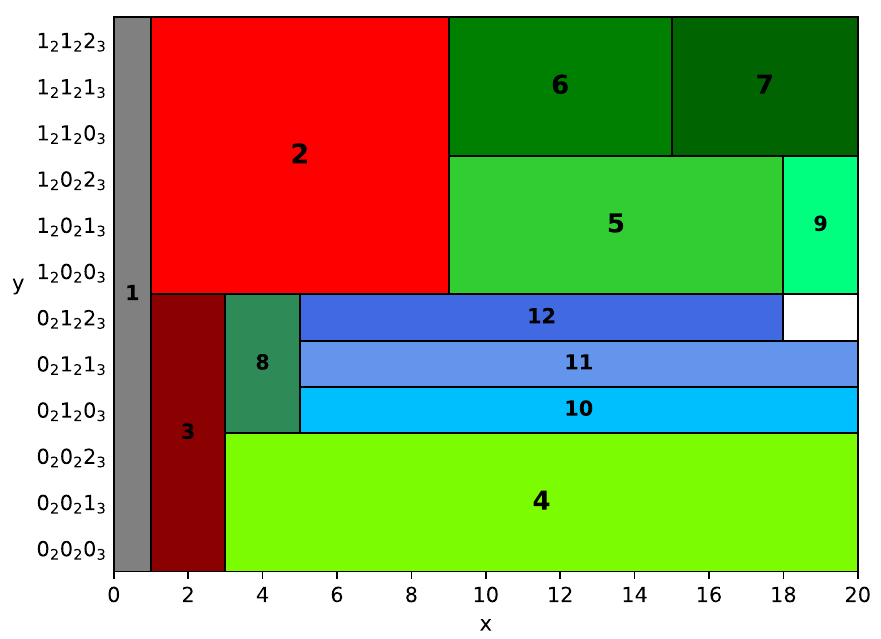}
   \caption{RG-FF-OPT}
   \label{fig:joffopt}
\end{subfigure}
\caption{Unfinished (a), (b) and finished (c) packings created by different methods on instance from Section~\ref{sec:equivalence}.}
\label{fig:example_solutions}
\end{figure*}

The optimistic variant fills the bags in a "preemptive manner" and always keeps at most a single bag not full. We start with an empty list of bags. Until we exhaust all the rectangles with height $H_{k+1}$, we select the widest $R_i$ with width $l_i$. If it fits any (i.e., the single not full) bag, we pack it to it. If the bag has vacant size $0 < A < l_i$, we split $R_i$ into two rectangles with widths $A$ and $l_i - A$. The first rectangle is packed into the bag, and the second is pushed back among the remaining rectangles. If all the bags are full, we create a new dummy rectangle of height $H_k$ and width $l_i$, push a new bag with size $l_i\cdot H_k/H_{k+1}$ to the list of bags, and pack $R_i$ into it; this new bag is a concatenation of $H_k/H_{k+1}$ bags of width $l_i$, that make up the rectangle. We denote this variant of the algorithm as \textbf{RG-FF-OPT}.

Note that only the last bag (and its corresponding dummy rectangle) might contain holes. In Figure~\ref{fig:opt_rectangles}, all the bags were actually fully filled. Also, notice that rectangles $R_5, R_7$ were split into two parts. Due to the splitting and the usage of a single rather than $H_1/H_2=2$ bags, we cannot simply partition the packed dummy rectangle into a feasible packing of rectangles that filled the associated bag; in a pessimistic case, it is possible (see Figure~\ref{fig:pess_rectangles}). However, since the dummy rectangles only provide a look-ahead, this "preemptiveness" does not cause issues with the final packing.

\subsubsection{First Fit Packing}
Once the dummy rectangles are constructed, our algorithm follows the first fit approach of algorithm \textbf{S-FF}. All the rectangles, both real and dummy, are sorted following the policy of Section~\ref{sec:heur_start}; first, non-increasingly by height, then non-increasingly by width. For feasibility, we need to ensure that the width $w$ of the 2D bin is not exceeded.

Our algorithm differs from \textbf{S-FF} in two following ways. Once the last rectangle of height $H_k$ is packed, we remove packed dummy rectangles from all the sub-bins, recalculate the sub-bins' utilizations, and then continue packing the rectangles of height $H_{k+1}$. The second difference is the packing policy. If the current rectangle $R_i$ fits any sub-bin, we pack it according to the first fit policy. If not, we do not immediately stop the algorithm; the sub-bins may become under-utilized once the dummy rectangles are removed. Thus, if $R_i$ is a dummy rectangle, we pack it to the least utilized sub-bin, temporarily exceeding the sub-bin's size. If $R_i$ is a real rectangle, we find the sub-bins that would fit $R_i$ after removal of dummy rectangles. If such sub-bins exist, we pack $R_i$ to the least utilized one; otherwise, the heuristic failed to find a feasible solution. If all real rectangles are packed, the algorithm finishes successfully, and we transform the packing into the periodic schedule.

Internally, we use the compressed bin tree representation of \cite{Eisenbrand2010} so the complexity of the first fit packing of a single rectangle is $\mathcal{O}(n)$ given $n$ rectangles to pack (i.e., jobs to schedule). Since for each height $H_k$ we need to process $O(n)$ real and dummy rectangles, the overall complexity of our solution is $\mathcal{O}(r\cdot n^2)$. 

\subsection{Comparison of Methods}
In Figures~\ref{fig:example_solutions}, we compare \textbf{T-FF}, \textbf{S-FF}, and \textbf{RG-FF-OPT} methods on example instance from Section~\ref{sec:equivalence}, using the $HD2D$ packing view. Neither \textbf{T-FF} nor \textbf{S-FF} methods were successful, as they were not able to reserve space for blue rectangle $R_{10}$. Also, notice the slight difference between these methods. \textbf{T-FF} puts $R_8$ to sub-bin $1_20_20_3$ - $1_20_22_3$, while \textbf{S-FF} uses sub-bin $0_21_20_3$ - $0_21_22_3$. While both methods process $R_8$ at the same moment, \textbf{T-FF} selects the time-wise earliest start time for the associated job, while \textbf{S-FF} selects the lowest sub-bin. 
Finally, \textbf{RG-FF-OPT} used dummy rectangles to reserve enough space for $R_{10}$ by putting rectangles $R_2, R_3$ into different sub-bins.

\begin{table*}[ht]
	\centering
\caption{Number of instances solved by each method for instance sets $S_1, S_2, S_3$.} 
\begin{adjustbox}{width=0.7\textwidth}
\begin{tabular}{||c | c || c  c  c  c  c  c || c  c ||}
	\hline
&  & \multicolumn{6}{c||}{heuristic} & \multicolumn{2}{c||}{exact}  \\
set & \# instances  & LPT & T-FF & S-FF & S-BF & RG-FF-PES & RG-FF-OPT & ILP & CP \\ \hline
$S_1$ & 3518 & 2213 & 3233 & 3248 & 3263 & 3378 & 3378 & 3518 & 3518\\
$S_2$ & 800 &  122 & 558 & 562 & 563 & 681 & 694 &  800 & 800\\
$S_3$ & 297 & 8 & 14 & 15 & 15  & 24 & 29 & 289 & 289 \\
 \hline
\end{tabular}
\end{adjustbox}
\label{tab:succ_methods}
\end{table*}

\section{\uppercase{Experiments}}
\label{sec:experiments}
The algorithms were implemented in Python 3.10. Experiments were performed on an Intel Xeon Silver 4110@2.10 GHz. IBM ILOG CP Optimizer v22.1.1 and Gurobi v10.0.2 were used as CP and ILP solvers, respectively. Both solvers used a single thread.

\subsection{Experimental Data}
We evaluated the implemented methods on several sets of synthetically generated problem instances. Due to the success of the method by \cite{hladik2020complexity}, we only generated feasible instances with utilization $\utilinst$ equal to 100 \%.

The first set of instances was generated using the scheme proposed in \cite{hladik2020complexity}. Starting with a selected period $\per{0}$, a single job with period and processing time equal to $\per{0}$ is generated. Then, for a given number of iterations, one of the jobs with period $\pertask{i} = \per{k}$ is selected and is either split into two new jobs of the same period but shorter processing times or is divided into $\base{k+1}$ jobs with the same processing time and period equal to $\per{k+1}$. Set $S_1$ contains original instances from \cite{hladik2020complexity}.

The original generation scheme often created instances with many long-period short-processing times jobs that are relatively easy to solve. We modified the generator so the jobs with shorter periods or longer processing times had a chance to be saved from splitting, thus creating instances potentially more complex to solve. Set $S_2$ refers to instances generated with the modified scheme, using the same periods as in the set $S_1$. Set $S_3$ used longer shortest period $T_0$ (80, 200, or 1000). While the sets $S_1, S_2, S_3$ contained only 80 jobs on average, their characteristics helped us compare the heuristics.

Finally, we generated 4 sets of difficult instances. They were generated in the canonical form of $HD2D$ packing and then transformed into the scheduling domain. We started with a single sub-bin of maximum height and divided it until sub-bins of height 1 were reached and filled with rectangles. We ensured that the majority of jobs had a processing time greater than 13. These short-jobs-free instances seemed to be the most difficult to solve; their sets were labeled by the letter $D$. Four sets with 200 instances were created. Each instance set $D_{T}^r$ used period set $\{800\cdot T^k~|~k \in 0,\dots,r-1\}$. Instances contained on average 84 jobs in case of $D_2^6$, 405 jobs in case of $D_3^6$, 3720 in case of $D_5^6$, and 1473 in case of $D_{20}^3$.

\begin{table}
	\centering
\caption{Number of instances solved by exact methods on difficult instance sets.} 
\begin{adjustbox}{width=0.25\textwidth}
\begin{tabular}{||c | c || c  c | c c||}
	\hline
set & \# instances &  ILP & CP \\
	\hline
$D_2^6$ & 200 & 95 &  152  \\
$D_3^6$ & 200 & 1 & 36\\
$D_5^6$ & 200 & 0 &  0  \\
$D_{20}^3$ & 200 & 0 & 0 \\
 \hline
\end{tabular}
\end{adjustbox}
\label{tab:succ_exact}
\end{table}

\begin{table}
	\centering
\caption{Number of instances solved by RG-FF-OPT and combinations of heuristics $M_i$ on instance sets $S_1,S_2,S_3$.} 
\begin{adjustbox}{width=0.45\textwidth}
\begin{tabular}{||c | c || c c c c c ||}
	\hline
set & \# instances & RG-FF-OPT & $M_1$ & $M_2$ & $M_3$ & $M_A$  \\  \hline
$S_1$ & 3518 & 3378 & 3423 & 3439 & 3437 & 3459\\
$S_2$ & 800 & 694 & 709 & 726 & 725 & 742\\
$S_3$ & 297 & 29 & 30 & 30 & 30 & 33 \\  
 \hline
\end{tabular}
\end{adjustbox}
\label{tab:succ_heur}
\end{table}

\begin{table*}[ht]
	\centering
\caption{Results of the utilization experiment on instance sets $S_1,S_2,S_3$.} 
\begin{adjustbox}{width=0.98\textwidth}
\begin{tabular}{||c | c || c c c c c c c c  c  c  c  c || c c c c ||}
	\hline
&  & \multicolumn{12}{c||}{heuristic} & \multicolumn{4}{c||}{exact}  \\
 & & \multicolumn{2}{c}{LPT} & \multicolumn{2}{c}{T-FF} & \multicolumn{2}{c}{S-FF} & \multicolumn{2}{c}{S-BF} & \multicolumn{2}{c}{RG-FF-PES} & \multicolumn{2}{c||}{RG-FF-OPT}& \multicolumn{2}{c}{CP10} & \multicolumn{2}{c||}{CP60}       \\ 
 set & \# instances & \# succ & avg $\utilinst_F$ & \# succ & avg $\utilinst_F$ & \# succ & avg $\utilinst_F$& \# succ & avg $\utilinst_F$ & \# succ & avg $\utilinst_F$ & \# succ & avg $\utilinst_F$ & \# succ & avg $\utilinst_F$ & \# succ & avg $\utilinst_F$ \\ \hline
 
$S_1$ & 3518 & 3452 & 0.962 & 3512 & 0.994 & 3512 & 0.995 & 3512 & 0.995 & 3518 & 0.998 & 3518 & 0.998 & 3518 & 0.999 & 3518 & 0.999   \\
$S_2$ & 800 & 788 & 0.923 & 798 & 0.988 & 798 & 0.989 & 798 & 0.989 & 800 & 0.997 & 800 & 0.996 & 800 & 1.000 & 800 & 1.000  \\
$S_3$ & 297& 277 & 0.915 & 296 & 0.964 & 296 & 0.965 & 296 & 0.966 & 297 & 0.978 & 297 & 0.977 & 297 & 1.000 & 297 & 1.000   \\
 \hline
\end{tabular}
\end{adjustbox}
\label{tab:util_methods}
\end{table*}

\subsection{Success Rates of the Methods}\label{sec:absolute_results}
First, we evaluated how many instances each method can successfully solve. CP and ILP solvers were limited to 3 minutes of computation time per instance. Heuristics were run once, as was outlined in Section~\ref{sec:heur_start}.

We found that the heuristics could not find any feasible solutions in the case of the difficult $D_T^r$ instances. Thus, we only used these instances to compare CP and ILP. As is reported in Table~\ref{tab:succ_methods} both solvers can solve smaller instances $S_1, S_2, S_3$. However, CP is more successful on the mentioned difficult instances shown in Table~\ref{tab:succ_exact}, given the limited computation time of 3 minutes. We suspect that the reason is an efficient formulation that relies on the powerful $\textbf{pack}$ constraint. However, for difficult sets $D_5^6,~D_{20}^3$, neither method found any solution. We use these sets later in Section~\ref{sec:utilziation_experiment} to highlight the shortcomings of the CP in comparison with heuristics.

The results of heuristics on sets $S_1, S_2, S_3$ are also reported in Table~\ref{tab:succ_methods}. We can see that on all these sets, the packing-inspired methods RG-FF-PES and RG-FF-OPT defeated the baseline heuristics, with the optimistic variant RG-FF-OPT being the best. This is most pronounced for set $S_2$. Our revised generation scheme made the instances harder to solve; this was highlighted by set $S_3$, where heuristics solved only a few instances. Finally, except for LPT, the results of baseline heuristics do not differ significantly.

We were also interested in the possible orthogonality of the studied heuristics; if each type of the heuristic would solve a distinct subset of the studied instances, we could run them as a portfolio and achieve a higher overall success rate while keeping the computation time short. To evaluate this, we combined the results of all different pairs of the heuristics, and we present several combinations. $M_1$ is combination of RG-FF-OPT and RG-FF-PES, $M_2$ is combination of RG-FF-OPT and S-BF, and $M_3$ is combination of RG-FF-OPT and T-FF. Finally, $M_A$ corresponds to running all the heuristics together.

From the results reported in Table~\ref{tab:succ_heur}, we can see that the combination of the two proposed methods does not offer much improvement. On the other hand, when we use our heuristic with the S-BF ($M_2$), the results are significantly better, increasing the number of successes by 30 in the case of set $S_2$. However, set $S_3$ remains problematic for heuristics to solve, even when we would run all the heuristics together ($M_A$). 

\subsection{Utilization Experiment}\label{sec:utilziation_experiment}
While the results reported in Section~\ref{sec:absolute_results} show the power of the CP, we also studied how each method tackles the instances with utilization $\utilinst < 1.0$. While the heuristics have trouble solving instances with $\utilinst=1.0$, they might be successful when $\utilinst$ is slightly lower. We study this in the following experiment.

For a given method and instance, we evaluate whether the method successfully solves the instance. If not, we remove the job with the smallest value of utilization $\utilization{i}$ and repeat the process. When the success eventually occurs, we report the final utilization $\utilinst_F$ of the instance. Thus, if the method solved each instance immediately or just after a few iterations, it would report all the values $\utilinst_F$ close to 1. Note that the jobs are removed in a deterministic way, as outlined; there might exist a situation when removing a single longer job would make the instance easier than removing several shorter jobs consecutively. We remove the jobs until the utilization of the instance reaches $\utilinst < 0.7$ when the failure is reported. The experiment was performed for each instance set, using all the heuristics, as well as the CP solver with a time limit: CP10 was limited by 10 seconds of computation, while CP60 was limited by 60 seconds.

\begin{figure}
\centering
\includegraphics[width=0.46\textwidth]{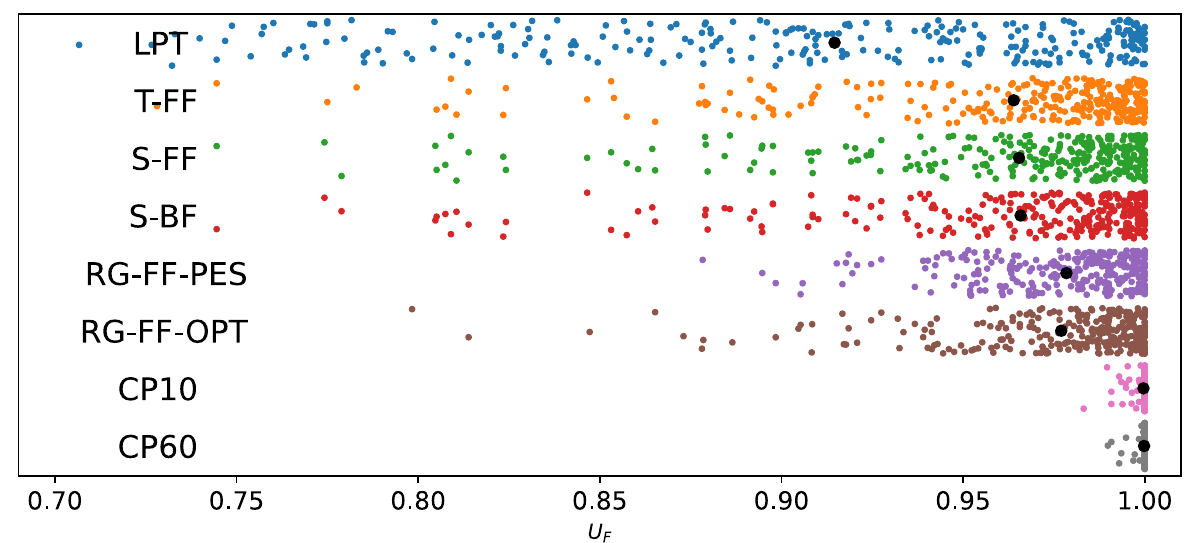}
\caption{Utilization experiment on instance set $S_3$. Colored dots mark final utilizations $\utilinst_F$ for each instance. Black dots mark the average value per method.}
\label{fig:utils3}
\end{figure}

\begin{figure}
\centering
\includegraphics[width=0.47\textwidth]{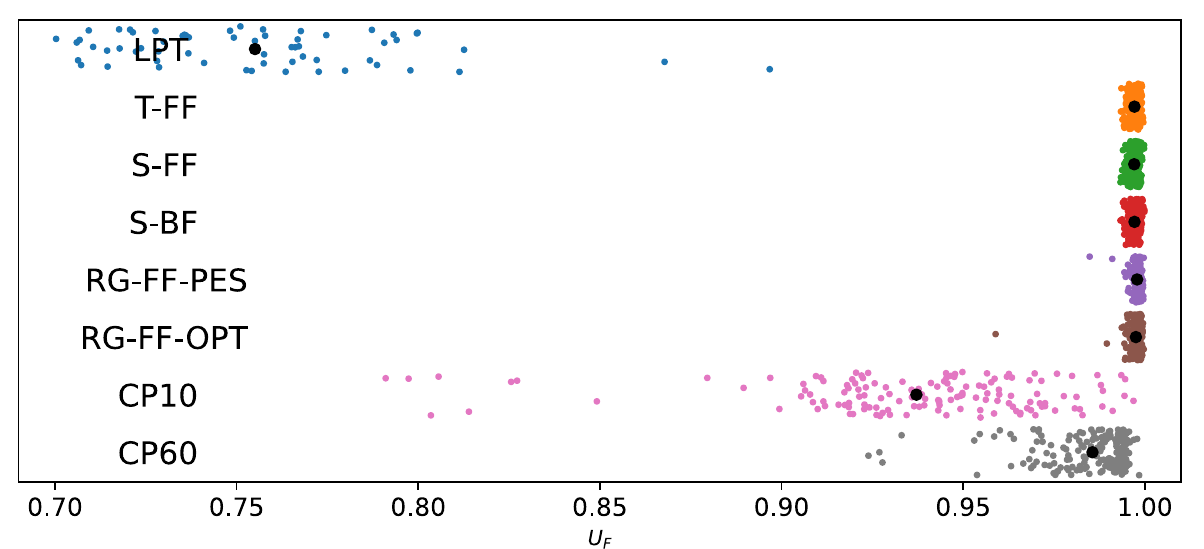}
\caption{Utilization experiment on instance set $D_5^6$. Colored dots mark final utilizations $\utilinst_F$ for each instance. Black dots mark the average value per method.}
\label{fig:utilh5}
\end{figure}

The number of successes (i.e., number of instances for which success occurred before $\utilinst_F < 0.7$) and an average value of $\utilinst_F$ of successful attempts on instance sets $S_1, S_2, S_3$ are shown in Table~\ref{tab:util_methods}. All the heuristics, with the exception of the LPT, are able to solve most of the instances with just a slight reduction of their overall utilization. This matches the results reported in \cite{hladik2020complexity}. On instance sets $S_2, S_3$ (see Figure~\ref{fig:utils3}), which were generated with our modified scheme, the proposed heuristics RG-FF-PES and RG-FF-OPT report average value of $\utilinst_F$ 1 \% higher than achieved by their baseline counterparts (with the theoretical maximum being 3 \%). These results again suggest that the proposed methods perform better on the instances with infrequent long jobs.

\begin{table*}[ht]
	\centering
\caption{Results of the utilization experiment on difficult instance sets.} 
\begin{adjustbox}{width=0.98\textwidth}
\begin{tabular}{||c | c || c c c c c c c c  c  c  c  c || c c c c ||}
	\hline
&  & \multicolumn{12}{c||}{heuristic} & \multicolumn{4}{c||}{exact}  \\
 & & \multicolumn{2}{c}{LPT} & \multicolumn{2}{c}{T-FF} & \multicolumn{2}{c}{S-FF} & \multicolumn{2}{c}{S-BF} & \multicolumn{2}{c}{RG-FF-PES} & \multicolumn{2}{c||}{RG-FF-OPT}& \multicolumn{2}{c}{CP10} & \multicolumn{2}{c||}{CP60}       \\ 
 set & \# instances & \# succ & avg $\utilinst_F$ & \# succ & avg $\utilinst_F$ & \# succ & avg $\utilinst_F$& \# succ & avg $\utilinst_F$ & \# succ & avg $\utilinst_F$ & \# succ & avg $\utilinst_F$ & \# succ & avg $\utilinst_F$ & \# succ & avg $\utilinst_F$ \\ \hline
$D_2^6$ & 200 & 188 & 0.873 & 200 & 0.995 & 200 & 0.994 & 200 & 0.995 & 200 & 0.994 & 200 & 0.992 & 200 & 1.000 & 200 & 1.000 \\
$D_3^6$ & 200 & 139 & 0.805 & 200 & 0.996 & 200 & 0.996 & 200 & 0.996 & 200 & 0.997 & 200 & 0.996 & 200 & 0.999 & 200 & 1.000 \\
$D_5^6$ & 200 & 61 & 0.755 & 200 & 0.997 & 200 & 0.997 & 200 & 0.997 & 200 & 0.998 & 200 & 0.998 & 144 & 0.937 & 200 & 0.986   \\
$D_{20}^3$ & 200 &  0 & - & 200 & 0.997 & 200 & 0.997 & 200 & 0.997 & 200 & 0.997 & 200 & 0.996 & 200 & 0.991 & 200 & 0.997 \\
 \hline
\end{tabular}
\end{adjustbox}
\label{tab:util_methods_difficult}
\end{table*}

Results on difficult sets $D_T^r$ are reported in Table~\ref{tab:util_methods_difficult}. Interestingly, most heuristics are able to solve them once the instances' utilization slightly drops.
Regarding the performance of the CP solver, interesting results are reported for the largest instances of $D_5^6, D_{20}^3$. There, the performance of CP10 and CP60 diminishes in contrast with heuristics (whose runtime is at most 0.2 seconds), as is also shown in Figure~\ref{fig:utilh5}. Thus, as the instance size grows and with a limited computation time, it may be better to use heuristics to create a slightly less utilized schedule.

\section{Conclusion}
\label{sec:conclusion}
In this paper, we studied the $PSP$ with a harmonic set of periods and non-preemptive jobs on a single machine. We formally proved that the studied problem is equivalent to the $HD2D$ packing problem and that each feasible schedule and its associated packing can be transformed into the canonical form. We used the CP with a bin formulation and showed its effectiveness compared to the ILP. Then, we proposed heuristics based on the $HD2D$ packing perspective, suitable for scheduling instances with infrequent long jobs. 

We have shown the advantages of our method in comparison with several baseline heuristics on sets of synthetically generated instances. Furthermore, we performed an experiment where we incrementally lowered the utilization of instances. The experiments allowed us to compare the heuristics and observe the CP's shortcomings when limited computation time is in place. The experiments have shown that the short-jobs-free instances with 100 \% utilization are the most difficult to solve. However, rate-monotonic-based heuristics work very well whenever the utilization slightly drops. The empty space of 99\% utilization instance behaves as infrequent, very short jobs, which can fill the fragmented space left at the end of the constructive heuristics run. 

We aim to use the presented results and heuristics as a basis of evolutionary algorithms to tackle more complex problems with dedicated machines, precedence relations, and a criterion to optimize.



\bibliographystyle{apalike}
{\small
\bibliography{bibliography}}
\end{document}